\documentclass{llncs}
\usepackage{makeidx}  
\usepackage{url}
\usepackage{graphicx}
\usepackage{float}
\usepackage{amsmath}
\usepackage{cleveref}
\usepackage{booktabs}
\usepackage[usenames, dvipsnames]{color}

\newcommand{\remove}[1]{}
\def\bit{%
	\leavevmode
	\vtop{\offinterlineskip 
		\setbox0=\hbox{B}%
		\setbox2=\hbox to\wd0{\hfil\hskip-.03em
			\vrule height .3ex width .15ex\hskip .08em
			\vrule height .3ex width .15ex\hfil}
		\vbox{\copy2\box0}\box2}}

\DeclareMathOperator{\dprime}{\prime \prime}
\begin{document}

\mainmatter              
\title{A New Look at the Refund Mechanism in the Bitcoin Payment Protocol}
\subtitle{(Full Version\footnote{This paper has been accepted to Financial Cryptography and Data Security 2018.})}

\author{%
Sepideh Avizheh\inst{1}  \and
Reihaneh Safavi-Naini\inst{1} \and
Siamak F. Shahandashti \inst{2}
}%
\institute{
University of Calgary, Calgary, Alberta, Canada\\
\email{(sepideh.avizheh1,rei)@ucalgary.ca}\\
\and
University of York, York, United Kingdom\\
\email{siamak.shahandashti@york.ac.uk}\\
}

\maketitle
%

\begin{abstract}
 BIP70 is the Bitcoin payment protocol for communication between a merchant and a pseudonymous customer.
 McCorry et al. (FC~2016) showed that BIP70 is prone to refund attacks and proposed a fix that requires the customer to sign their refund request. 
 They argued that this minimal change will provide resistance against refund attacks.
In this paper, we point out the drawbacks of McCorry et al.'s fix and propose a new approach for protection against refund attacks using the Bitcoin multisignature mechanism.
 Our solution  does not rely on merchants storing refund requests, and unlike the previous solution, allows  updating refund addresses through email. We discuss the security of our proposed method and compare it with the previous solution. 
We also propose a novel application of our refund mechanism in providing anonymity for payments between a payer and payee in which merchants act as mixing servers.
We finally discuss how to combine the above two mechanisms in a single payment protocol to have an anonymous payment protocol secure against refund attacks.
\end{abstract}

\section{Introduction}
Since the introduction of Bitcoin in 2008~\cite{BitcoinSatoshi}, it has been widely adopted by merchants as a payment method. 
By 2015, the number of merchants accepting Bitcoin was reported to surpass 100,000~\cite{IBT} and it has continued to expand into new markets (see e.g.~\cite{Bitcoin-Japan,Bitcoin-Korea}). 
Bitcoin standards are developed through a process which involves a so-called (Standard Track) Bitcoin Improvements Proposal (BIP) being proposed, discussed, ratified, and adopted by the Bitcoin community.
BIP70~\cite{BIP70} is  the Bitcoin Payment Protocol standard that defines the communications between a pseudonymous customer and a  merchant with a public key certificate. 
 The protocol, provides a number of properties to improve the interaction between the two entities (e.g.\ allows the merchant's address  to be  human-readable) and also provides the necessary guarantees (e.g. a proof of payment to the customer that can be used for dispute resolution).
 One important feature of the protocol is that the customer  can specify refund addresses that will be used by the merchant
  in the case of refunds for cancelled orders or overpayments.

 McCorry, Shahandashti and Hao however showed
two {\em refund attacks} on the BIP70 protocol~\cite{mccorry}. These attacks are referred to as the Silkroad Trader and Marketplace Trader attacks, and exploit the inadequacies of the refund mechanism in the standard, including for example the fact that the protocol only provides one-way authentication.
 In the Silkroad Trader attack, a malicious customer uses the refund mechanism to relay a payment to an illicit merchant (the Silkroad Trader) through an honest merchant. This is by simply declaring the Bitcoin address of the Silkroad Trader as the refund address, and later asking for a refund. McCorry et al.~\cite{mccorry} discuss the steps required for the attack in detail and showed its feasibility by successfully carrying out the attack in real-life payment scenarios. 
 The authors also described a second attack,  called Marketplace Trader attack, in which  a rogue trader
plays the role of a Man-in-the-Middle (MITM)  
 between the customer and a reputable merchant  and effectively direct the customer's payment to its own address after using the inadequate authentication of the protocol to change the refund address.
 In both these attacks,  the analysis of blockchain data will not reveal  the attacks.
 In McCorry et al.'s solution the customer signs the refund addresses by the public key they have used in the payment (the signature is called a \emph{proof of endorsement}) and so
effectively binds the refund addresses to the customer address. This prevents the first attack since the customer cannot deny their link to the
Silkroad trader anymore. McCorry et al.~\cite{mccorry} argue that this measure also discourages the second attack since merchants will become more reluctant to update the refund address through unauthenticated channels such as email. 

 McCorry et al.'s solution, although  minimally changing the protocol, introduces a major data management challenge for the merchant. This is because protection against  the Silkroad trader attack 
 requires 
the merchant to maintain a database of  
the proofs of endorsement and transactions that are signed by the customer. 
	These transactions include the refund   information  including the amount and the refundee address that the merchant must use, and the amount and address of the customer that the merchant has received the bitcoins from. Using the stored data, the merchant can ``prove" that they have followed the customer's request and have  not colluded with the customer in  transferring money to the refundee (Silkroad Trader).  We refer to this as an {\em explicit log solution} where all the relevant information must be stored and kept indefinitely by the merchant. The stored data are also privacy sensitive and so the merchant must adopt extra measures to secure the storage.  
 The data must be kept in the database indefinitely as  proof may become necessary at any time in the future. 
The database must  be securely backed up  (e.g. using cloud services) to ensure data is available when required. 

\vspace{1mm}
\noindent
{\bf Our contributions.}
 { We introduce  {\em implict logging}  that 
 requires the merchant to only store a number of indexes to the blockchain  that will be used to recover the required proof of innocence, 
 when needed. 
 The solution works as follows: a refund operation consists of two transactions produced by the merchant. The first transaction is a two output transaction with the refund amount that requires the signatures of the refundee and the customer to release the fund. The second transaction has the same value and is issued to the customer, 
 with a time lock that allows the transaction to be released only after a specified  time passes.
The refundee requires the signature of the customer to receive their refund, which implies that they endorse the refund. If the customer does not endorse the refund (e.g. in the case of a Marketplace Trader attack), the first transaction will not be redeemed in time and hence the customer will be able to use the second (time-locked) transaction to redeem the bitcoins.

 The merchant will store the indexes of the two transactions that are issued for a refund request, together with the indexes of the original payment transaction and the redeemed  transaction.
 In the case of 
 multiple refundees,  all their details will be included in the first transaction and so the same amount of storage will be required.
  
 The solution is robust  in the sense that transactions  that are associated with a refund and  contain the proof of the relation between the customer and the refundee are kept on the blockchain and  are  immutable. 
 If the database that contains the indexes of the transactions is corrupted, 
 the merchant can still recover the information about the transactions that they have issued or received, by searching the blockchain 
for transactions that include their public key information. 
 The solution preserves the privacy of the refund transaction against  an adversary who has access to the blockchain  
  and eavesdrops on the communication 
  between the customer and the merchant, in the sense that the linkage between  the customer and the refundee can only be revealed by the customer or the merchant.}  
%
  We will discuss how the scheme can be modified
  when there are  more than one transaction issuers (this 
   was considered by McCorry et al.).
  The details of this  solution is given in  \Cref{New}.
  
 We show that this solution also provides protection against 
 Marketplace Trader attack without 
 putting any restriction on BIP70. 
 This is in contrast with  McCorry et al.'s  solution 
 that requires the refund address 
not be accepted through email.  This restriction 
is likely to be ignored in practice, rendering McCorry et al.'s solution icapable of preventing Marketplace Trader attacks. 
  
In \cref{Anonymity} we consider a novel application of the refund mechanism in providing payment anonymity.  The main observation is that refund addresses in Bitcoin can effectively provide a level of indirection that if carefully used, can decouple the payer and payee.
In our proposal, the merchant provides a mixing service that allows the customers to pay for other services and to other merchants using a combination of overpayment and refund.  By  ``mixing" transactions of  multiple customers, the linkage of transactions using their payer and payee fields, as well as values, will be removed.  We  define the communication protocol between the customer and the mixing service based on BIP70.
 Finally, in \Cref{Aggregate} we discuss how the above two mechanisms can be combined to provide an anonymous payment protocol with security against refund attacks.

\section{Preliminaries}\label{Preliminaries}
\subsection{The BIP70 Payment Protocol}
BIP70 \cite{BIP70} is a Bitcoin application layer payment protocol that defines the  sequence  of messages communicated between a customer and a merchant. 
BIP70 consists of three messages: \textit{payment request}, \textit{payment}, and \textit{payment acknowledgment}. It proceeds as follows.


After the customer selects an item from the merchant's website and clicks to pay, the merchant responds by sending a \textit{payment request} message. This message contains \textit{payment details}, the  information related to merchant's X.509 certificate (PKI type and PKI data), as well as the signature of the merchant on the hash of the payment request.
Here \textit{payment details} consists of the Bitcoin address that the customer should send the bitcoins to, the time that request has been created, an expiration time, a memo containing notes to the customer, a payment URL, and finally the merchant data which is used by the merchant to identify the payment request.

The customer's Bitcoin wallet subsequently verifies the signature and the merchant's identity, the information in the payment details, such as the time of the request creation and expiry, displays the merchant's identity, the amount to pay, and the memo to the customer and asks the customer whether they want to continue.
If confirmed, the wallet will create the necessary Bitcoin transactions for the payment and broadcast them to the Bitcoin peer-to-peer (P2P) network.
Then, a \textit{payment} message is sent to  the merchant. This message consists of the merchant data from the \textit{payment details} in payment request, one or more valid Bitcoin transactions, the \textit{refund$\_$to} field which specifies a set of refund amount and address pairs
to be used in case of a refund request, and a note for the merchant (memo).


When the merchant receives the payment message, it verifies that the transactions satisfy the payment conditions, broadcasts the transactions, and sends back a \textit{payment acknowledgment} message. This message contains a copy of the payment message and a final memo including a note on the status of the transaction.


BIP70 does not specify how the payment request message should be downloaded, but requires that the payment and payment acknowledgment messages are communicated over a secure channel (such as HTTPS).

BIP70 does not explicitly define a refund protocol. It is implicitly assumed that if the customer requests a refund identifying the payment by the \textit{merchant data} field,
the merchant issues a refund transaction which sends the refund amounts to the corresponding refund addresses specified in the \textit{refund$\_$to} field of the \textit{payment} message.

\begin{figure}[t]
\small
\centering
\setlength{\belowcaptionskip}{-14pt}
\includegraphics[width=\textwidth]{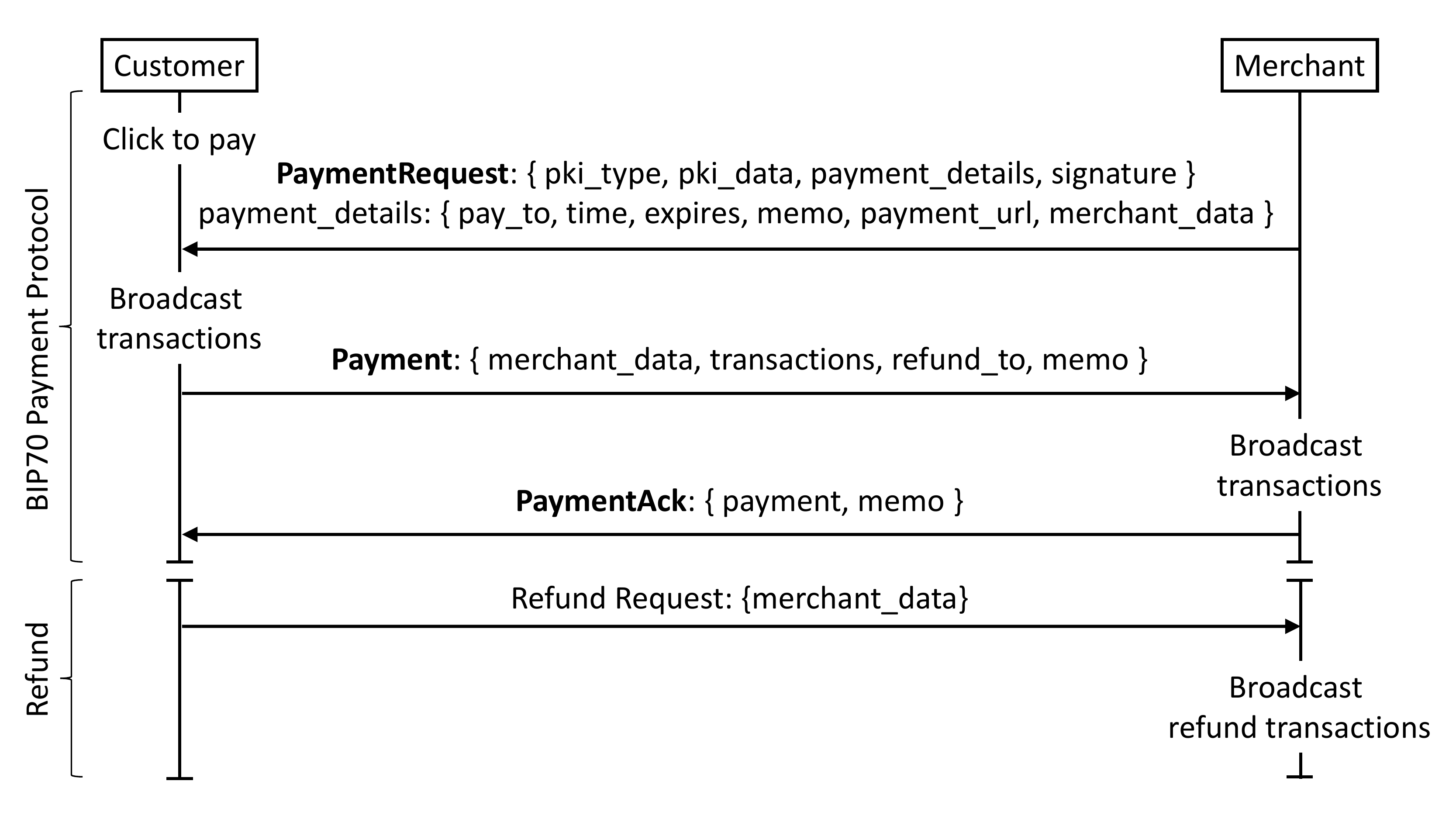}
\caption{The BIP70 payment protocol and its refund procedure.
         Note that the Bitcoin P2P network to which the transactions are broadcast is not explicitly shown here.}
\label{F:BIP70}
\end{figure}

\Cref{F:BIP70} shows the communication flow in BIP70 and its implicit refund procedure. 
Note that besides communicating with each other, both the customer and the merchant are assumed have access to the Bitcoin P2P  network. Both are able to broadcast transactions to the Bitcoin P2P network. Upon receiving a transaction, the Bitcoin P2P network decides whether to add the transaction into the distributed ledger (i.e. the Bitcoin blockchain) through the Bitcoin consensus mechanism. 
Both the customer and the merchant are also able to check whether their broadcasted transaction has been included in the blockchain. 
It is important to distinguish the communications in the application layer payment protocol and those in the P2P network. To simplify our diagrams, we do not explicitly show the P2P network communications. 

\subsection{Refund Attacks}
McCorry et al. propose two attacks on the refund process of BIP70~\cite{mccorry}.
These attacks work even if a secure channel such as HTTPS is used for communication between parties.
We briefly describe these two attacks in the following.

\begin{figure}[t]
\small
\centering
\setlength{\belowcaptionskip}{-14pt}
\includegraphics[width=\textwidth]{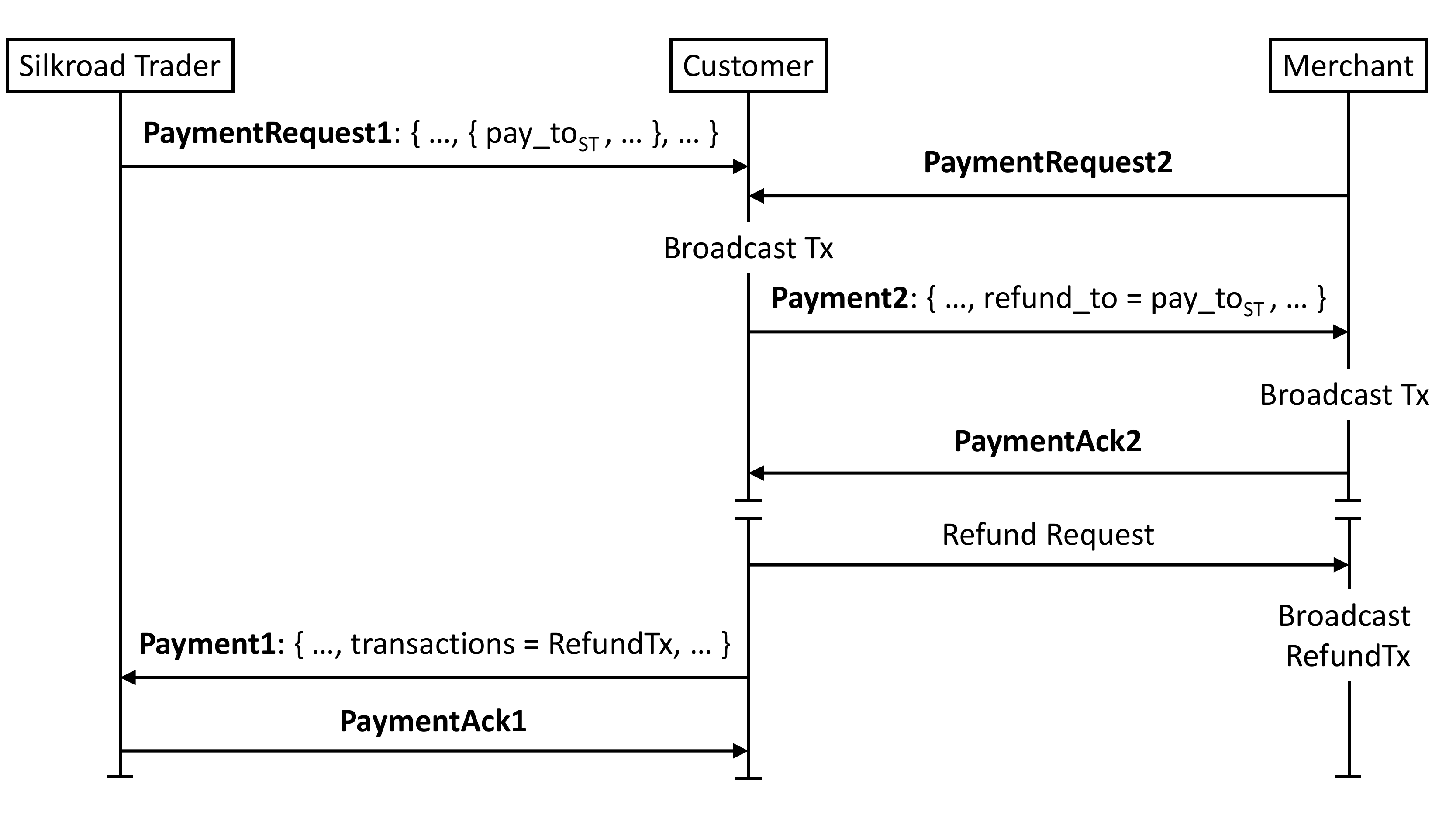}
\caption{The Silkroad Trader attack.}
\label{F:Silkroad}
\end{figure}

\vspace{1mm}
\noindent
{\bf Silkroad Trader attack.}
The refund addresses provided by the customer (in the \textit{refund$\_$to} field) are in no way endorsed and can be repudiated at a later time. This means that a malicious customer may abuse the refund mechanism to relay their payment to an illicit trader (here called the Silkroad trader) through an honest merchant. The customer simply provides the illicit trader's address as the refund address to the merchant and thus when a refund is requested, the merchant will send the refund to the illicit Trader. The customer can later deny abusing the refund mechanism and the merchant will have no way to prove they have been cheated.
\Cref{F:Silkroad} shows the interaction among parties in this attack.

\begin{figure}[t]
\small
\centering
\setlength{\belowcaptionskip}{-14pt}
\setlength{\abovecaptionskip}{-4pt}
\includegraphics[width=\textwidth]{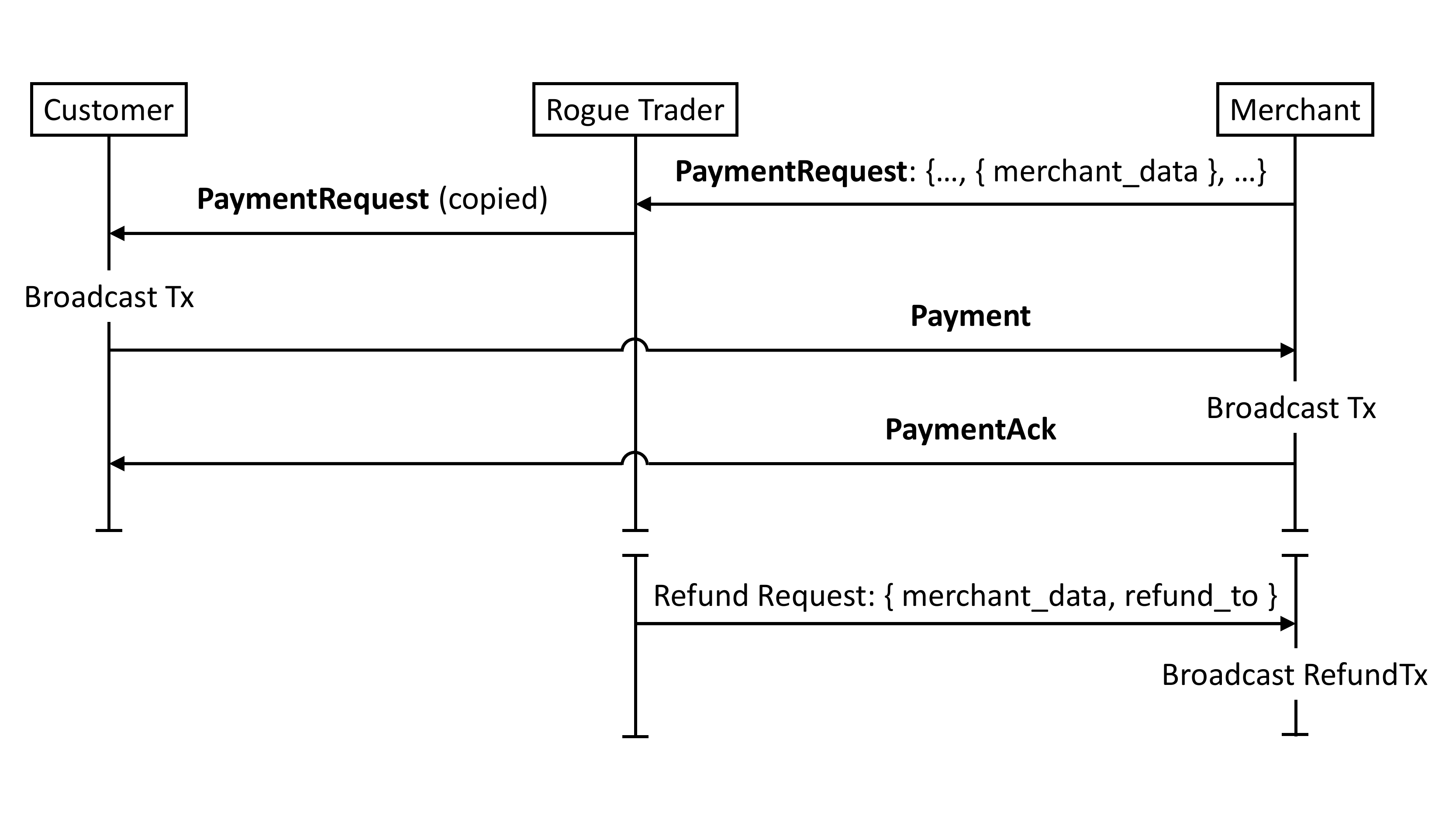}
\caption{The Marketplace Trader attack.}
\label{F:Marketplace}
\end{figure}

\vspace{1mm}
\noindent
{\bf Marketplace Trader attack.}
Some merchants allow customers to specify new refund addresses upon a refund request. The customer requesting the refund is not authenticated. This means that any entity who has knowledge of the payment identifier (specified in the \textit{merchant data} field of the payment details in the payment request message) can request a refund to any arbitrary account. This is the basis for the Marketplace Trader attack, in which a rogue trader acts as relaying man-in-the-middle for the payment request message between the merchant and the customer. Hence, the rogue trader is able to find out \textit{merchant data}. At a later time, the rogue trader requests a refund to an arbitrary address and is able to steal the funds.
\Cref{F:Marketplace} shows the interactions among the parties in this attack.

\subsection{McCorry et al.'s Solution to Refund Attacks}
McCorry et al. propose to include in the payment message a ``proof of endorsement'' for refund addresses.
To do this, each customer address involved in the payment protocol is required to produce a digital signature on (and therefore ``endorse'') a corresponding refund address.
Employing this solution, at the end of a successful payment protocol, the merchant will be in possession of a proof of endorsement for each refund address.
Such a proof can be presented and verified by a third party in case of a Silkroad Trader attack to implicate the malicious customer.
Besides, since such a proof of endorsement is valuable for merchants, McCorry et al.\ argue that it will discourage merchants to accept new refund addresses unless accompanied by a proof of endorsement, resulting in reducing the possibility of Marketplace Trader attacks.


In McCorry et al.'s solution, for guaranteed protection against attacks, merchants will need to store payment transactions as well as payment requests and payment messages which are required to verify the proof of endorsement.
Therefore, the actual storage overhead of McCorry et al.'s solution is much larger than only keeping proofs of endorsement. As noted earlier, maintaining a secure and robust database to store proof of endorsement messages is a security bottleneck of the system and can particularly become expensive for smaller merchants with limited resources.
\remove{McCorry et al.\ acknowledge other limitations of their solution due to Bitcoin inherent problems, including transaction malleability, an idiosynchrasy of Bitcoin.  
In particular, the customer can tamper with the transaction data as a whole (by slightly changing the signature script or re-signing the transaction) and then re-broadcast it to the network. Such tampering (e.g.\ adding dummy data to the signature script or re-signing) will change the transaction hash and hence the proof stored in the database will no longer match the transaction hash, rendering the stored proof ineffective. Transaction malleability is not an issue in some updated versions of Bitcoin, but the original protocol is still in use. }
\subsection{Multisignature and Time-Locked Transactions in Bitcoin}
Although it is convenient to think of Bitcoin transactions as sending funds to certain account addresses, technically what the transaction specifies is a set of redemption criteria in a certain script language. Any subsequent transaction which satisfies the redemption criteria may authorize the transfer of funds made available in the original transaction.

The most popular script is ``Pay to Public Key Hash'' (P2PKH), which requires a signature corresponding to an address, hence effectively sending the bitcoins to the address. Typical Bitcoin transactions use this script.

Another popular and more versatile script is ``Pay to Script Hash'' (P2SH), which requires satisfying a script, the hash of which is listed. P2SH can be used to implement a diverse range of transactions including \emph{multisignature} transactions. A $k$-of-$n$ multisignature transaction requires $k$ signatures corresponding to $k$ addresses within a set of $n$ specified addresses to be present to redeem the funds in the transaction.

An interesting script which can be combined with the ones discussed above is one that effectively freezes the transaction funds until a time in the future to create a so-called \emph{time-locked} transaction. The funds in a time-locked transaction cannot be spent by any other transaction until a certain (absolute or relative) time in the future.

\section{A New Approach to Protection against Refund Attacks}\label{New}

We propose a solution to refund attacks that  
requires the merchant to store a fixed number of indexes (and so constant size) in each run of BIP70 protocol.
The solution is robust to possible damages to database content, 
and ensures privacy of refund transaction from outsiders.
 
BIP70 requires the \emph{payment} and \emph{payment acknowledgment} to be sent over a secure channel, however does not specify such a requirement for the \emph{payment request} message \cite{BIP70}. 
Therefore, in general we consider  two  types of attackers: 

\begin{itemize}
\item \textbf{Online attacker}, intercepts the communication channel and sees all the input/output messages of a merchant; they also have access to blockchain data.
\item \textbf{Offline attacker}, has only access to the blockchain data.
\end{itemize}

To simplify our description,  we first assume BIP70 communication is over HTTPS and so we only need to consider an offline attacker.
 We then show how to secure the protocol against an online attacker.

Our goal is to provide the following properties for the refund mechanism: 
 \begin{description}
{ \item[\textbf{Implicit log.}] The merchant only stores  indexes of  transactions in a protocol run. 
 This has the 
  following advantages:
 	\begin{itemize}
 		\item \textbf{Constant storage size per protocol run. } 
		 The local storage size for a 
		 customer's payment  in a protocol run  
		  is constant.  
 		\item \textbf{Robustness.} The refund mechanism will work correctly and  reliably in the case of  
		a dispute, even if the merchant's local database is corrupted or lost. 
		In the worst case when none of the locally stored indexes are accessible, the merchant can recover the required proofs by searching the blockchain using their own public key information. 

 	\end{itemize} 
 
\item[\textbf{ Refund privacy.}]   An 
offline adversary with access to the blockchain, or 
an online adversary with access to 
the blockchain and the communication link between the merchant and the customer, cannot 
reveal the linkage between the customer and the refundee.
Note that BIP70 does not require secure communication between the customer and the merchant and so an online adversary can access unencrypted communication between the two.
We note that the merchant's local database must be kept secure for customers' privacy. }
 \end{description}

We also aim to conform with BIP70 specifications and avoid extra restrictions including {\em not} accepting refund addresses by email.
 Note that Refund addresses are valid for two months from the time of the payment \cite{BIP70}, and during this
 period
  the customer should be allowed  to change the refund addresses
  for example when an existing refundee %
  has lost
  their wallet.
  Coinbase and Bitpay \cite{Coinbase,BitPay}  both
  accept refund address updates via email.

 \subsection{Our Solution}\label{Solution}
\begin{figure}[t]
\small
\centering
\setlength{\belowcaptionskip}{-14pt}
\includegraphics[width=90mm]{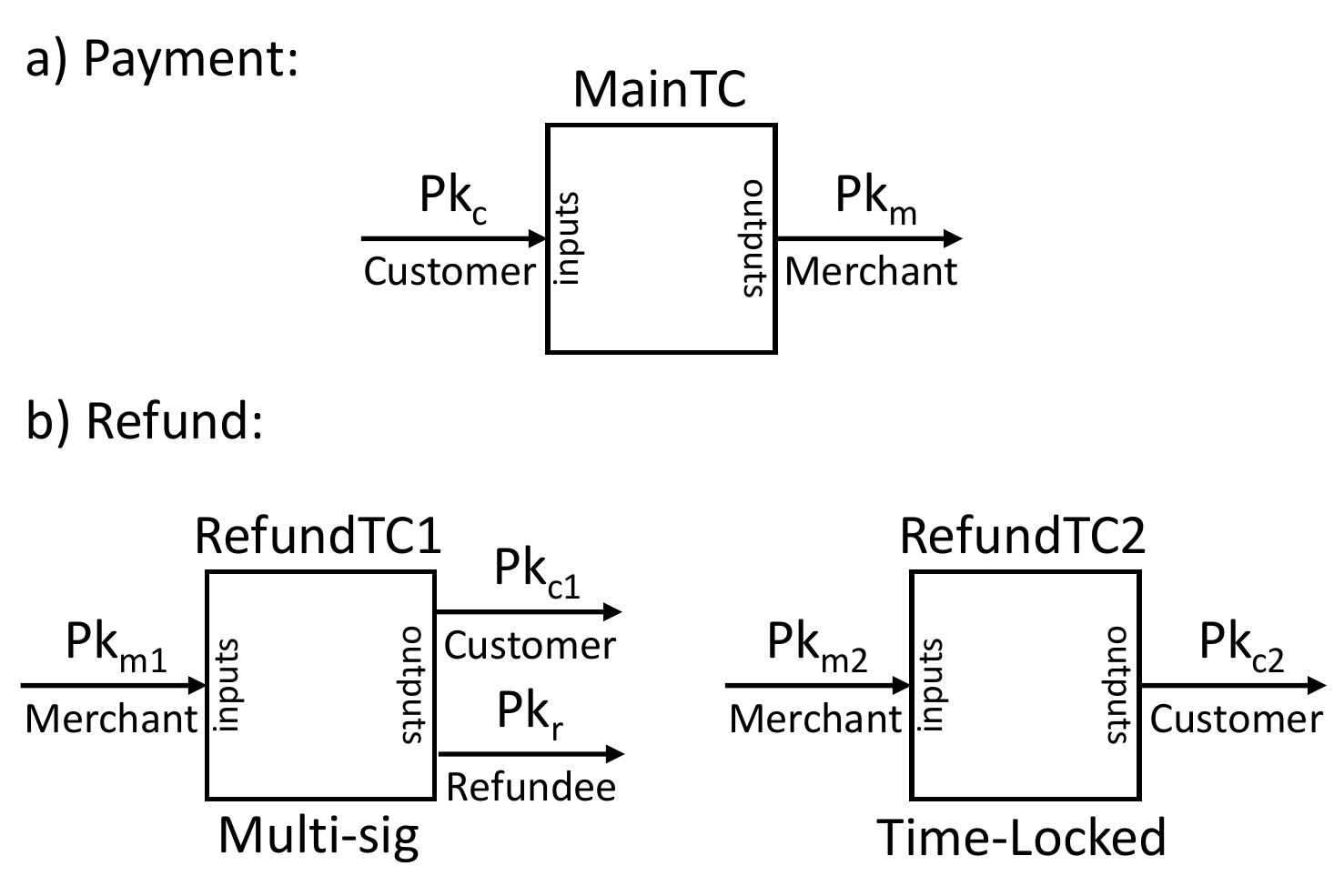}
\caption{(a) The main transaction. (b) The proposed Refund mechanism, in which merchant locks the transaction to the customer and the refundee and they may redeem this transaction if they collaborate. The merchant also issues RefundTC2 for robustness to ensure that the refund can be unlocked by customer in case the first transaction is not redeemed.}
\label{F:SolutionTxs}
\end{figure}

\remove{
  {\color{red} Our approach substantially reduces 
  	the challenges of maintaining a database of customer's private information for an unlimited time period, in a secure and robust way.}
  We  use capabilities of Bitcoin transactions, namely multisignature and time-locked transactions, to link the refund addresses to the customer.
  Thus
  the merchant's ''evidence for innocence''
  is stored in an immutable  distributed database.
}
Our proposed  refund mechanism works as follows.
The merchant creates a 2-of-2 multisignature transaction, and hence binds the refund amount to both the customer and refundee.
Then, to make the protocol robust
in case one of the addresses is not available, a second transaction is created.
This second transaction is a time-locked transaction, and the customer is its only recipient.
Merchant uses a lock time for this transaction to give priority to the first transaction. If the customer and the refundee
could not collaborate to redeem the refund, the customer is able to claim them after the lock time.
Note that the lock time creates a delay in the system only if the
 customer does not know the refundee, e.g.\ in the case of a Marketplace Trader attack.
 In other words, the second transaction is
 a backup for  system robustness (see \Cref{F:SolutionTxs}).

  In addition,   { to preserve refund privacy 
  against offline attackers}\footnote{
  Bitcoin transactions use fresh addresses (address freshness)~\cite{Wiki} 
  to protect the privacy of the address owner as well as others. 
  }
 the merchant deterministically creates fresh addresses from the public key of the customer and then masks  them with a Diffie-Hellman key, generated
  using the fresh address of the customer and the private key of the merchant.
 This is to ensure that
 only  the merchant or the customer who are able to re-generate the Diffie-Hellman key can discover the linkage between the refund key and the payment transaction.
   To derive fresh addresses we 
   assume the customer has  a deterministic wallet based on BIP32 \cite{BIP32}.  
Most  Bitcoin wallets support BIP32 and  so this is a reasonable assumption. 
In the rest of the paper, by deterministic wallet we mean a BIP32 wallet with a master key and key hierarchy (keys can be specified with an index).

Based on BIP32, a deterministic wallet generates a tree of public/private key pairs on elliptic curve $E$, e.g.\ for a 1-level tree, it creates $2^{31}$ {\em hardened}  and $2^{31}$ {\em non-hardened } keys.
Hardened keys are  public keys for which the associated private keys can only be known before the generation of the public key.
Non-hardened keys however allow
anyone to derive a valid child public key from the parent public key,  while the owner of the
 parent (master) private key can generate the respective child private key.
  In our protocol, the customer address in the payment transaction is a non-hardened public key which is used as a parent key by the merchant to derive  child public keys. The customer knows the respective private keys and  can also create the Diffie-Hellman key using the child private key and the public key of the merchant. Hardened keys can be used for refund addresses. The 
  step by step process is given below:

\begin{description}

\item[\textbf{Key generation.}]The customer wallet software generates a tree of public/private key pairs using BIP32 \cite{BIP32}.
Each private key is an element of  $F_q$, where $q=p^n$ is a prime power.  Each public key is a point on an elliptic curve $E$  (specified for bitcoin) over $F_q$. $H_l$ denotes the 256 leftmost  bits of HMAC-SHA512 which is used for computing the child key; the rightmost bits of HMAC-SHA512 are used as the next level chain code.
Let
$Pk_c=cP$ be a non-hardened public key.  A child public key can be derived by anyone using this
 parent public key
 as follows:
 $Pk^{\prime}_{c}=Pk_c+H_{l}(ch_c,Pk_c||index)P$;
 where $Pk^{\prime}_{c}$ is a child public key, $Pk^{\prime}_{c}=c^{\prime}P$, but only the customer who knows the parent private key can compute the child private key $c^{\prime}$. $ch_c$ denotes the chain code which is the 256 rightmost side of the parent hash, and $index$ is the index of the generated child key in the tree.   $(Pk_c,ch_c)$ is called the {\em extended pubic key} (see \cite{BIP32} for details). 
 We also use $H^{*}$ to denote 
a collision resistant hash function which maps a point on Elliptic curve $E$ to $F_q$, this is used for computing the Diffie-Hellman component in child keys as we describe it below.

\item[\textbf{Click to pay.}] The customer visits the merchant website and chooses an item, then declares their intent to pay (e.g.\ by clicking a ``pay'' button). 
\item[\textbf{Payment request.}] The merchant sends the payment request including their public key, $Pk_m=mP$. This public key is unique for each transaction.
\item[\textbf{Payment message.}]  The customer after authenticating 
the merchant,  puts together  a payment transaction, 
\textit{MainTC}, that transfers the cost of the chosen item to the merchant,  and uses 
 a non-hardened extended public key  (with external chain) $(Pk_c=cP,ch_c)$, as a data output (using the OP$\_$Return opcode).  
Finally, the customer creates a \textit{Payment} message based on \textit{MainTC}, and
{specifies their extended public key}, $n$ refund addresses, $(Pk_{r_1},Pk_{r_2},$ $\dots,Pk_{r_n})$, and  the amount of refund for each address.
\item[\textbf{Payment ack.}] The merchant detects \textit{MainTC}, and returns a \textit{PaymentAck} message to the customer.
\item[\textbf{Refund request.}]  Within a 2 month period from the payment request \cite{BIP70}, the customer can use the addresses provided in \textit{Payment.refund$\_$to} field to receive their refund.  In this case, the merchant does the following:

\begin{enumerate}
\item { Generates two new keys, $Pk_{m_1}=m_1P$ and $Pk_{m_2}=m_2P$.}
\item Derives child keys from the customer extended public key as described earlier, $Pk^{\prime}_{c_i}$ for $1 \leq i \leq n+1$.
\item Masks  the child keys as $Pk^{\dprime}_{c_i}=Pk^{\prime}_{c_i}+H^{*}(m_{1}Pk^{\prime}_{c_i})P$ for $1 \leq i \leq n$, and $Pk^{\dprime}_{c_{i}}=Pk^{\prime}_{c_{i}}+H^{*}(m_{2}Pk^{\prime}_{c_{i}})P$ for $i=n+1$. 
\item  { Creates and broadcasts two transactions, RefundTC1 (in which the merchant's public key is $Pk_{m1}$) and RefundTC2 (in which the merchant's public key is $Pk_{m2}$). RefundTC1 is a P2SH transaction that can be redeemed by providing signatures from both $Pk^{\dprime}_{c_i}$ and  $Pk_{r_i}$ for $1 \leq i \leq n$. RefundTC2 is a P2PKH transaction that can be redeemed by the private key corresponding to $Pk^{\dprime}_{c_{n+1}}$ after a specified
 lock time period (e.g.\ one week).}
\end{enumerate}

\remove{
\begin{figure}[h]
\small
\centering
\setlength{\belowcaptionskip}{-14pt}
\includegraphics[width=80mm]{Solution1.jpg}
\caption{Protection against Refund attacks; Communication flow is similar to BIP70 \cite{BIP70}. Upon a refund request, the merchant creates two transactions to lock the bitcoins to both the customer and the refundee. The merchant also generates a transaction which pays the same amount to the customer only; this latter transaction has a lock time.}
\label{F:SolutionFlow}
\end{figure}
}

\end{description}

\subsection{Protection against Silkroad Trader Attacks}
In the Silkroad Trader attack, the customer wants to   remove their link
to the Silkroad Trader  using  a victim merchant.
In our approach
the refund transaction can be only redeemed if the customer and the refundee collaborate. 
Hence, the redemption of the refund transaction constitutes an evidence of the linkage between the customer and the refundee.
This linkage, however,  is hidden from those who observe the Bitcoin blockchain
and so the merchant is the only one who knows this linkage.
The merchant can prove that a payment transaction and a refund transaction are linked to each other by
first deriving the child keys from the payment transactions and adding the respective Diffie-Hellman key
(based on the public key of the merchant and the customer's child key) to them, then using the refundees' public key and the customer's masked child keys to create the P2SH address and show that it matches the address within the refund transaction; finally showing that the refund transaction has been redeemed by a pair of collaborating customer and  refundee, thus establishing the evidence of the linkage.
The
merchant does not need to store the transactions in their database and the proofs are robustly preserved on the blockchain.

\subsection{Protection against Marketplace Trader Attacks}
If the customer provides a refund address during the BIP70 protocol run and later update it via email, the  merchant just uses the newest refund address and locks the refund amount to both the refundee and the customer.
 If the customer and the refundee collaborate to redeem the refund, the refund is finalized; otherwise, if the rogue trader sends their own address to the merchant, they cannot later claim the bitcoins since the customer will not collaborate with an unknown refundee to sign the refund transaction. 
 In case of such an attack, 
the customer will be able to claim the refund 
after the lock time expires. 
Thus, the approach  protects the
customer against Marketplace Trader attacks and at the same time provides
the  possibility to update the  refund addresses.

\subsection{Communication over HTTP}
{
	BIP70 does not restrict the customer and merchant to use HTTPS. Our proposed solution to refund attacks although by design hide the relation between the customer and the merchant, \textit{payment} messages include  information such as the refundee's address which can be used by an online attacker to find the corresponding refund transactions by searching the blockchain for the transactions that contain the address and so trace refundees transactions. 
	To protect against these attacks, the customer may use  HTTPS, or 
	Diffie-Hellman key agreement on the  merchant's public key and their own private key, to generate a key that will be used  to encrypt sensitive messages such that they are decryptable by the merchant.  Alternatively, merchant can follow the key generation algorithm of our scheme for both customers and refundees (for refundees derive child keys of the refund addresses and then mask them with Diffie-Hellman key on the refundees child key and their own private key). In this case refundee can still find the corresponding private key to claim the refund, but attacker cannot detect the refund transactions.}

\subsection{Analysis}
In the following, we show how each of the mentioned properties will be satisfied.

\textbf{Implicit logging.}
A merchant may store   information  that are communicated during a payment protocol for various reasons, including  bookkeeping,  refund or exchange, or statistics about  customers and products.
Here we do not  consider bookkeeping that is mainly for accounting purposes or the  ability to honour refund or exchange policies.   Nor we consider data storage that are for statistical analysis purposes.
As noted in \cite{mccorry}, using the Bitcoin Payment Protocol requires the merchant to store evidence to protect them against refund attacks. This information must be kept for a sufficiently long time to be effective in providing such protection. 

Consider a single run of the payment protocol. Transactions which are the ``evidence for innocence'' (of the merchant) are stored on the blockchain. The merchant must just store the transaction indexes, that is, the transaction ids (txid) of  the \textit{MainTC}, both the refund transactions  \textit{RefundTC1} and \textit{RefundTC2}, and \textit{Redeem transaction} by which the customer (and possibly the refundee) redeem the refund. Table~\ref{tab1} shows a full refund record in the merchant's database. 
Each transaction index is  32 bytes and so $4\times32=128$ bytes are needed for the four transactions ids (\textit{MainTC}, \textit{RefundTC1}, \textit{RefundTC2}, and \textit{Redeem transaction}).
Note  that  the merchant can always use a deterministic approach for the child key indexes, for example always start from index 0 and increment it for each new child key, and so
they do not need to be stored. It is not difficult to see that storage size is independent of the
number of refundees.

\begin{table}
	\setlength{\abovecaptionskip}{-8pt}
	\caption{One refund record in the merchant's database.}
	\label{tab1}
	\centering
	\footnotesize
	\scalebox{0.9}{
		\begin{tabular}{c @{\qquad} c @{\qquad} c @{\qquad} c @{\qquad} c }
			\hline 
			MainTC txid & RefundTC1 txid & RefundTC2 txid & Redeem txid \\
			\hline
			32 bytes  & 32 bytes & 32 bytes & 32 bytes \\
			\hline 
	\end{tabular}}
	\vspace{-5mm}
\end{table}

In the case of a dispute,
the merchant can retrieve all the mentioned transactions from the  blockchain using their stored transaction ids,
and then use the chain code, index, and the public key of the customer, to derive the related child key, and
use their
own private key $m_i$, $i=1,2$, to re-create the masked address, $Pk^{\dprime}_{c}=Pk^{\prime}_{c}+H^{*}(m_iPk^{\prime}_{c})P$.
If the output of the P2PKH transaction with address $Pk^{\dprime}_{c}$ is a spent output, it indicates that the customer has redeemed the refund without the collaboration of the refundee. If the redeemed transaction is the P2SH transaction, $Pk^{\dprime}_c$ and the refund addresses in the \textit{Redeem transaction} are used to re-generate the P2SH address. If the address matches the \textit{RefundTC2} and it is a spent output we know that the owner of the masked child public key $Pk^{\dprime}_{c}$ (or more precisely the owner of the public key $Pk_c$) and refundee have collaborated with each other to redeem the refund. 

\textbf{Minimal storage size.} In the following we compare the storage cost of our proposed protocol with that of McCorry et al.~\cite{mccorry}. The comparison summary can be found in Table~\ref{tab}.
In McCorry et al.'s solution, the proof of endorsement  is a signature that must be locally stored.
Verification of this signature requires information about the main transaction and the communicated messages  including the refund addresses, refund values, the  memo from the customer, and the payment request message. The merchant also needs to store the transaction ids for both the payment and refund transactions.
Let $L_S$ denote the size of the proof of endorsement signature. 
The size of a transaction input with one signer is at least  146 bytes {\footnote{Previous transaction hash  is 32 bytes, previous Tx-out index is 4 bytes, Tx-in script length is 1-9 bytes, public key is 33 bytes in compressed format, signature is 72 bytes, sequence number is 4 bytes.}}.  Other values are, refund address which is 34 bytes, refund value which is 8 bytes, memo and payment request message sizes are denoted by
$L_{pay}$ and   can  reach 50,000 bytes. Finally  a transaction id is 32 bytes.  Thus in total,
for one refundee, $252+L_S+L_{pay}$  bytes must be stored at the merchant side and this cost grows linearly   with the number of refundees.
The total storage for $n$ refundees will be
$210+42n+L_S+L_{pay}$ bytes which is significantly higher than our scheme.
Note that we are not considering the size of the merchants' keys in our calculations.

\begin{table}[h]
	\vspace{-1em}
	\centering
	\footnotesize
	\setlength{\abovecaptionskip}{-6pt}
	\caption{Storage size (in bytes) of our approach vs. \cite{mccorry}}
	\label{tab}
	\scalebox{0.9}{
		\begin{tabular}{l@{\qquad}c@{\qquad}c}
			\toprule
			Scenario & McCorry et al.~\cite{mccorry}& Our approach\\
			\midrule
			1 refundee  & $252+L_S+L_{pay}$ & 128\\
			$n$ refundees & $210+42n+L_S+L_{pay}$ & 128\\
			\bottomrule
	\end{tabular}}
	\vspace{-5mm}
	\label{tab2}
\end{table}

\textbf{Robustness.}
The payment protocol must work correctly  in case of a dispute or when the information stored in the merchant's database is corrupted or lost.
In \cite{mccorry},  if the local database that stores the signature  (proof of endorsement) is corrupted,   the evidence of the collusion will be irreversibly lost and the merchant will become vulnerable to refund attacks.
In our  proposed approach however the merchant can exhaustively  search on all their keys to retrieve the database records. To do so, the merchant re-generates all the private/public keys using their wallet (through the master key) and then uses a blockchain explorer (e.g. blockchain.info) to search for the transactions that contain these public keys. We denote the search complexity on blockchain with $O_S$. Assuming the number of keys used by the merchant to be $2^k$, the search complexity becomes $2^kO_S$. The retrieved transactions are then identified as \textit{MainTC}, \textit{RefundTC1}, and \textit{RefundTC2} based on their types and the merchant role as a sender or receiver of each transaction. If the merchant is the recipient, the transaction is \textit{MainTC}. If the merchant is the sender and the transaction is P2SH, it is \textit{RefundTC1}, otherwise if the merchant is the sender and the transaction is P2PKH, the transaction is \textit{RefundTC2}. \textit{Redeem transactions} are also found by searching the output addresses in \textit{RefundTC1} and \textit{Refund TC2}. If the number of these transactions is $\ell$ the search complexity increases to $2^k.\ell.O_S$. After this classification, the merchant can follow the steps below to retrieve the database:
\begin{enumerate}
	\item The merchant chooses one specific \textit{MainTC} transaction and stores its index in the database.
	\item To find the related refund transactions, the merchant should reconstruct the masked child key of the customer $Pk^{\dprime}_c$. First, merchant generates the child key of the customer, $Pk^{\prime}_c$, using the chain code stored in \textit{MainTC} (for the index, merchant can always try the first index, i.e.\ pick index 0 and 1). Then, they mask the child key with the Diffie-Hellman component, $H(m_iPk^{\prime}_c)P$ to generate $Pk^{\dprime}_{c}$. For this, the merchant should try all the private keys $m_i$, for $ 1\leq i \leq 2^k$. 
	\begin{itemize}
		\item If $Pk^{\dprime}_{c}$ matches the key inside \textit{RefundTC2}, the used index is stored in the database as \textit{RefundTC2} txid. If this transaction is a spent transaction, merchant stores the corresponding \textit{Redeem transaction} index in the database and halts, since the proof is fully retrieved and customer have spent the transaction alone. 
		\item Otherwise, $RefundTC1$ may have been spent, so the merchant uses $Pk^{\dprime}_{c}$ to find the corresponding \textit{Redeem transaction} and uses the refund keys in that transaction to reconstruct the P2SH address.  The corresponding \textit{RefundTC1} is the one which contains the mentioned P2SH address. The index of this transaction is stored in the database and the proof is fully retrieved. 
		\item However, if \textit{Redeem transaction} for a specific customer does not exist at all, it shows that customer has not redeemed the refund value yet, hence the merchant should wait for the customer to claim the refund and subsequently retrieve the database record.
	\end{itemize}
\end{enumerate}

The complexity of reconstruction depends on the number of keys the merchant has used, the key generation, the search operation, and the number of refund transactions. If key generation (calculating the point arithmetic and hashing related to masked child key) takes $O_K$, and the total number of customers is $t$, then the total complexity is upper bounded by $2^k (2 t O_K+ \ell O_S)$. The factor $2$ for $O_K$ is because we should follow this procedure for both refund transactions \textit{RefundTC1} and \textit{RefundTC2}.

\textbf{Privacy.} To show that our scheme guarantees privacy, that is, only the merchant and the customer can reveal the linkage between the customer and the refundee, we consider two types of attackers: online attackers and offline attackers. Since an online attacker has access to both the communication channel and the blockchain, and hence is stronger, we only provide the justification for privacy for an online attacker. Privacy against offline attackers who only have access to the blockchain is straightforward. 

Suppose that the channel is not TLS-protected (worst case), so the online attacker can intercept all the communicated messages between the customer and the merchant. Through \textit{PaymentRequest} and \textit{Payment} message, the attacker discovers the public key of the merchant $Pk^{*}_m$, the extended public key of the customer  $(ch^{*}_c,Pk^{*}_c)$ 
, and the public key of the refundee $Pk^{*}_r$. The goal of the attacker is to link the customer address $Pk^{*}_c$ in MainTC to the refund address $Pk^{*}_r$. For simplicity we assume that \textit{RefundTC1} and \textit{Redeem transaction} is also given to the attacker; this simulates the case where the refundee is the attacker. In practice this assumption may not be true, and the attacker needs to also guess the transactions.
Overall, the attacker observes the information given in Table~\ref{view}.

\begin{table}
	\setlength{\abovecaptionskip}{-6pt}
	\caption{View of an online attacker.}
	\label{view}
	\centering
	\footnotesize
	\scalebox{0.9}{
		\begin{tabular}{ l @{\qquad} c @{\qquad} l }
			\hline
			Information & key & source(s) \\
			\hline
			Extended public key of customer  & $(ch^{*}_c,Pk^{*}_{c})$ & MainTC \\
			Public key of refundee  & $Pk^{*}_{r}$ & Payment message, Redeem transaction \\
			Public key of merchant  & $Pk^{*}_{m_1}$ & RefundTC1 \\
			Public key of customer  & $Pk^{\dprime}_{c}$ & Redeem transaction \\
			\hline
	\end{tabular}}
	\vspace{-5mm}
\end{table}

From \textit{Redeem transaction}, the attacker can link the customer address $Pk^{\dprime}_{c}$ to $Pk^{*}_r$. The reason is that to claim the bitcoins both the customer and the refundee need to sign the \textit{Redeem transaction}. Hence, a spent refund transaction shows that the customer with $Pk^{\dprime}_{c}$ knows the refundee with  $Pk^{*}_r$. Thus, to link the customer $Pk^{*}_c$ to refundee $Pk^{*}_r$, the attacker needs to just find whether $Pk^{\dprime}_{c}$ is derived from $Pk^{*}_{c}$. 

$Pk^{\dprime}_{c}$ is generated as follows $Pk^{\dprime}_{c}=Pk^{\prime}_{c}+H^{*}(m^{*}_1c^{\prime}P)P$, where $Pk^{\prime}_{c}$ is the child key derived from the customer address $Pk^{\prime}_{c}=Pk^{*}_{c}+H_{l}(ch^{*}_c,Pk^{*}_{c}||index)P$. The attacker knows $Pk^{*}_{c}$ and $ch^{*}_c$, so they can guess $Pk^{\prime}_{c}$ by trying different indexes with probability $\frac{1}{n}$, assuming the number of refundees is $n$. To link the child key $Pk^{\prime}_{c}$ to $Pk^{\dprime}_{c}$, the attacker needs to solve the decisional Diffie-Hellman problem (DDH) given $m^{*}_1c^{\prime}P$, $Pk^{*}_{m_1}=m^{*}_1P$, and $Pk^{\prime}_c=c^{\prime}P$. Since solving DDH is hard (say with $\epsilon$ representing the probability of solving DDH), the attacker's probability of success will be $P_{success}=\frac{\epsilon}{n}$ which is negligible.

\subsection{Multi-Signer Payment Transaction}
\remove{
So far we assumed that payment transaction is generated by a single customer. In the following
 we show that
 a functionality  similar to  \cite{mccorry},  for multiple signer case can be provided.
}
 In multi-signer transactions, MainTC has  multiple 
inputs (multiple public keys and signatures). 
This type of 
 transactions is generated for example when a number of 
 bitcoin addresses jointly pay for a transaction. 
\remove{
These kinds of transactions are generated in group payments when several entities intend to buy an item and chip in to pay, or when one entity collects the bitcoins received from different transactions, for example a customer has received 1\bit from each of his previous transactions while the item costs 5\bit. The public key of these transactions are different from each other due to address freshness rule in Bitcoin.}
When a payment transaction  is created by multiple signers,
 endorsing  the refund address by a single signer has the danger of allowing them  to steal the bitcoins or carrying out a Silkroad Trader attack. McCorry et al.'s solution is resilient against this attack because each key that is used in the payment transaction endorses its own refund address.
 In our scheme however, the customer does not provide any signature before the refund, hence the merchant does not know which refund address belongs to which signer in \textit{MainTC} and blindly locks the bitcoins to all of the signers. Although this solution prevents refund attacks, it is not efficient in the sense that the refundee must
   interact with all of the signers (that they may not know)
   to claim the refund.
   We use the \textit{refund$\_$to} field in the payment message; 
   if there are $\ell$ customers with public keys $Pk_{c_1}$, $Pk_{c_2}$,...,$Pk_{c_\ell}$ and $n$ refund addresses as $Pk_{r_1}, Pk_{r_2},...,Pk_{r_n}$, the \textit{refund$\_$to} field will be $\{(Pk_{c_1},Pk_{r_1},v_1), (Pk_{c_2},Pk_{r_2},v_2), ...\}$. This binding is authenticated later in our protocol, {meaning that if customer approves to sign the redeem transaction, the refundee's address is correct}, and so no signature is needed 
   at this stage. 

 We review possible attacks after introducing this modification.
 In the Silkroad Trader attack, the customer
 may modify the \textit{Payment.refund$\_$to} field and insert the Silkroad Trader address as a refund address of another signer. 
 Since the merchant will lock the refund transaction to the victim co-signer,
 the refund cannot be claimed by the Silkroad Trader
 since the victim co-signer does not collaborate with the unknown Silkraod Trader.
 The co-signer can claim the refund on their own through the second transaction (i.e.\ the P2SH transaction) issued for them by the merchant.
  In the Marketplace Trader attack, a co-signer may intend to change the refund addresses after the payment is finalized to steal the bitcoins. In this case, the co-signer will present a new refund address for each key used in the payment transaction. Again the merchant locks the bitcoins to the main customer and the new refundee and so
   the bitcoins will stay locked
    since the attacker cannot obtain the signature of the  customer on that transaction.
    Furthermore, the customer can redeem the refund from the second issued transaction (i.e.\ the P2SH transaction).
Note that 
 \begin{enumerate}
 \item A co-signer cannot change the value of refund through email. If the value is changed, the merchant will lock the bitcoins to all of the customers to ensure that they know about the change.
 \item When the payment transaction is a multi-signature transaction, each of the signers who has authorized the payment transaction must include at least one refund address. Even if two parties agree on one refund address, both must include it and the merchant must lock the refund to both of them through a 3-of-3 multi-signature output (or n-of-n if the number of them is $n-1$). If a signer fails to include a refund address, the Silkroad Trader attack becomes probable.
 \end{enumerate}

\remove{
	For  added  efficiency  of our scheme, we can add a flag in the payment message sent from customer to merchant to indicate whether the customer wants to have the second transaction (the P2PKH) for redeeming the bitcoins or not. This option is to reduce the burden on merchant and it is a sensitive information that must be sent in an encrypted format. Whenever the customer uses the refund address that he knows, he can choose this option, but note that in this case the update ability of refund addresses through email should be disabled to impede the Marketplace trader attack.
}

\section{Bitcoin User Anonymity Using Refund Mechanism}
\label{Anonymity}
Despite using pseudonym for senders and receivers of transaction, it has been shown that transactions can be linked \cite{Androulaki,Reid,Ron} and combined with other data possibly reveal user identities.
There have been a number of approaches for providing anonymity \cite{StealthAddress,CoinJoin,CoinSwap,FairExchange}.
Stealth address schemes \cite{StealthAddress} guarantee address anonymity against an online attacker who intercepts the communication link and sees the Bitcoin address of the payee when it is sent to payer to create the transaction.
 By stealth address technique, payer adds a Diffie-Hellman key to the payee's address in a way that the corresponding  private key is still known by payee. In CoinSwap \cite{CoinSwap} a party uses an intermediary node to send the payment to payee; the goal is anonymity against online attacker. In CoinJoin \cite{CoinJoin} a number of  parties agree to create one transaction together, they also use values with equal worth to provide value anonymity against an offline attacker. In Fair Exchange \cite{FairExchange} two people exchange their bitcoins with each other to achieve coins with a history that is unrelated to them, to resist against an offline attacker. Each of these solutions can be considered as a traditional mixing service. Users can also mix their coins through a mix server (e.g. bitmixer.io \cite{BitMixer}), which receives their coins and pay them back a fresh coin, although mixer receives a fee from the user. This technique, however has a problem, user should trust the mix server that they  will not steal their money.

Refund mechanism  provides a level of indirection that can be  used for adding anonymity to Bitcoin users.
We propose to use
 merchants as a trusted mixing servers by using a modified
  BIP70 protocol refund's policy. To use this service, the customer visits the website of a merchant and selects an item
  for purchase.
  By using overpayment
  and the recipients' addresses as the refund addresses, the sender can send payment to refundees in an anonymous way.
  Alternatively, they can send the  desired amount to the merchant and  later cancel their  order for the refund.
  In these situations, merchant acts as an intermediary  to allow the customer to pay the bitcoins to the recipients indirectly.
  Merchant can  also split the value to smaller chunks and mix the refund transactions of different entities to provide value and time anonymity respectively.
Reputable merchants are generally  trusted and are expected to follow the protocol.
Note that the merchant does not know if  the refundee in a
 refund transaction is a customer and
 cannot relate the output bitcoin addresses to the user.
 Merchants can benefit for  such service  by requesting a fee for it. 
%

In our proposed protocol, the merchant receives inputs from customers' transaction, and issues transaction with outputs based on the addresses in the \textit{Payment.refund$\_$to} field. { Refund addresses are extended public keys of refundees.} Merchant generates child keys of the respective refund addresses, split the values of refund to smaller equal chunks and sends the partitioned values to child keys. Merchant considers the refund address as a parent key and uses its child keys for refund transactions to have a fresh address for each chunk.
To provide confidentiality for  the parent refund key (this is needed because we assume online attacker exists and the communication is HTTP), the  merchant encrypts the child keys using the Diffie-Hellman key generated by the public key of the merchant and  the private key of the customer.
For secure mixing, the time relationship between  the input and the output of the mix must be  protected.
Otherwise an adversary who intercepts the merchant's channel can link
the two using the time information of the merchant input and output.
In the following protocol the merchant   mixes the refunds of different customers and  hides the time relation between inputs and outputs of the mix service.

\begin{description}
\item[\textbf{Key generation.}]We assume that Refundee is using BIP32 \cite{BIP32} wallets, and sends their extended public key to the customer. Customer generates a private/public key pair as $Pk_c=cP$.
\item[\textbf{Click to pay.}] The customer visits the merchant website and chooses an item, then clicks on ``pay''.
\item[\textbf{Payment request.}] The merchant sends the payment request message including their public key, $Pk_m=mP$. This public key is unique for each transaction.
\item[\textbf{Payment message.}] After authenticating the merchant, the customer picks a public key, $Pk_c$, and generates \textit{MainTC} which pays the cost of the chosen item to Merchant. Then, the customer creates a payment message with extended refund keys, $(Pk_{r_{i}},ch_{r_i}), \forall 1 \leq i \leq n$, for $n$ refund addresses, and the amount of refund value for each; then encrypts the \textit{Payment.refund$\_$to} field using Diffie-Hellman key $H^{*}(cPk_m)$ {(this  is not required if channel is TLS-protected).}
\item[\textbf{Payment ack.}] The merchant detects \textit{MainTC}, decrypts the refund addresses, and returns an acknowledgement message, \textit{PaymentAck}, to customer.
\item[\textbf{Refund request.}]  Within a predetermined distance from payment request (can be defined in \textit{Payment.refund$\_$to} field), customer can use the addresses provided in \textit{refund$\_$to} field to receive the refund.  In this case, the merchant

\begin{enumerate}
\item Splits each refund value (for different customers) to $k$ partition; for example, $v_1$ is divided to $v_{11}$, $v_{12}$, $\dots$ $v_{1k}$, and $v_2$ to $v_{21}$, $v_{22}$, $\dots$ $v_{2k}$ and so on. The goal is to have equal amounts.
\item Derives child keys of each refund address, $Pk^{\prime}_{r_{ij}}$ $\forall 1 \leq i \leq n$, and $\forall 1 \leq j \leq k$.
\item Creates a few transactions. For each child key merchant creates an output that pays its chunk of refund value to the corresponding masked child key $Pk^{\dprime}_{r_{ij}}=Pk^{\prime}_{r_{ij}}+H^{*}(m^{*}Pk^{\prime}_{r_{ij}})$, $\forall 1 \leq i \leq n$, and $\forall 1 \leq j \leq k$. While mixing the outputs of different customers in each transaction, the merchant broadcasts the transactions to the Bictoin network.
\end{enumerate}
\end{description}

\remove{
\begin{figure}[h]
\small
\centering
\setlength{\belowcaptionskip}{-14pt}
\includegraphics[width=80mm]{Splitting.jpg}
\caption{(a) Splitting the value for each refundee, (b) Deriving child keys of each refund address.}
\label{F:SolutionAnonymity}
\end{figure}}

\subsection{Discussion}
 Our approach is relatively similar to CoinJoin \cite{CoinJoin}. CoinJoin has two versions; in the first version users should agree and join to create one transaction. Because this co-joined transaction mixes inputs and outputs of different users, an offline attacker can not distinguish the relation between them, provided that the inputs are all in the same range. In the second version of CoinJoin, users trust a third party who mixes the input and outputs of different users and create one transaction, and sends it back to each user to sign. In this procedure, the third party learns the relation between inputs and outputs and IP addresses of the users. This approach cannot protect against online attackers who intercept the communication channels. To hide the IP address from online attackers, Tor or VPN must be used \cite{CoinJoin}.

\begin{figure}[h]
\small
\centering
\setlength{\belowcaptionskip}{-14pt}
\includegraphics[width=100mm,height=40mm]{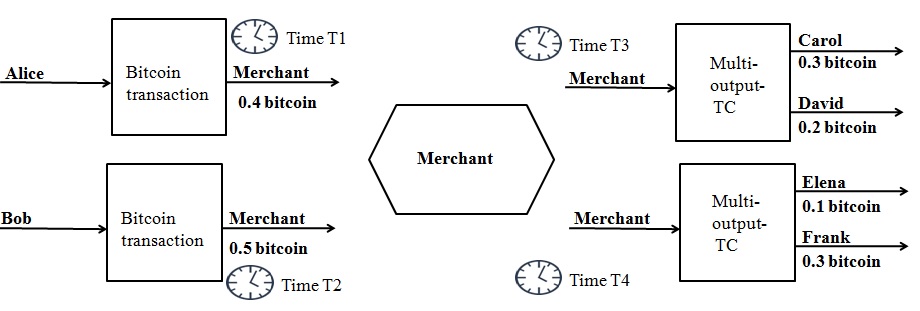}
\caption{Alice wants to send 0.4\bit to Carol and Elena, and Bob 0.5\bit to David and Frank. Alice pays the bitcoins to merchant, and introduces Carol and Elena as refundees, through BIP70. Bob also pays the bitcoins to merchant and introduces David as refundee. Merchant creates a few refund transactions and mixed their outputs and send them to refundees.}
\label{F:Comparison}
\end{figure}

 In our scheme, customers start a normal purchase from a merchant, and send the excess transferred amount, or alternatively the whole payment, to the refundee through merchant. First of all, it should be noted that refund request can be sent via a different communication channel such as email (by providing payment acknowledgment), so an online attacker is not able to see the refund request. Second, an online attacker can never be assured that the customers have really made a purchase or not, or even which item they have bought, since it is possible to make overpayments to hide the exact amount of payment. Third, in CoinJoin the final transaction consists of output  addresses who have been introduced by the owner of input addresses, meaning that everyone knows that the output address has a relation with one of the inputs (possibly ownership), and even the values in the input and output can leak information about the links between input and output addresses. 
 %
 In our scheme the merchant creates the refund transaction using a different address from the one used  in the MainTC, and so 
 the link between 
 the customer's first transaction (MainTC) and the refund transaction (RefundTC) is  removed. Additionally, any relation between the inputs and outputs in the sense of time, value, and address are taken off and a global passive attacker who monitors communication links and the blockchain cannot obtain any information (see \Cref{F:Comparison}). 
 Fourth, merchants will not be able to steal the bitcoins of the customers because of their reputation. This feature is necessary for any indirect payment; in CoinSwap \cite{CoinSwap}, which is also an indirect payment method, first a commitment transaction is created to impede theft and then the transfer is done. But in our scheme even if the merchant does not deliver the bitcoins to the refundee, the customer can present their payment acknowledgement message to a judge and prove the theft. Effectively, using the merchant as a mixing server extends the search space for linking transactions and ensures a higher anonymity compared to CoinJoin.

\section{Concluding Remarks}\label{Conclusion}
We proposed a new approach to mitigate Refund attacks against BIP70 using 
implicit logging  which requires the merchant 
to only store   indexes of four transactions for each run of the protocol.
Our approach provides a  solution that is robust against possible corruption of the merchant's local database, and preserves the privacy of the refund (i.e.\ the link between the customer and the refundee).
We also showed that refund mechanism can be used to provide 
anonymity for  payers and payees, through merchant acting as a mix server, 
and provided the communication protocol between the customer and the mixing service based on BIP70.
%
{ This is a novel approach for providing anonymity for bitcoin transaction that need careful evaluation. This will be an interesting  direction for future work.} 

%
%

\bibliographystyle{splncs03}
\bibliography{sample}

\appendix

\section{An anonymous BIP70 secure against Refund attacks} \label{Aggregate}

In this section, we aggregate the schemes in previous sections to have a unique scheme which is resilient against Refund attacks \cite{mccorry} and an offline attacker against anonymity of users.

\begin{description}

\item[\textbf{Key generation.}]Customer and refundees wallet software generates a tree of public/private key pairs using BIP32 \cite{BIP32}.
\item[\textbf{Click to pay.}] Customer visits the merchant website and chooses an item, then clicks on ''pay".
\item[\textbf{Payment request.}] Merchant sends the payment request message including their public key, $Pk_m=mP$. This public key is unique for each transaction.
\item[\textbf{Payment message.}] After authenticating and authorizing the merchant, customer chooses one of the non-hardened extended public keys, $(Pk_c,ch_c)$, and generates \textit{MainTC}; this transaction sends the cost of the chosen item to Merchant. The chain code of customer is also stored as a data in \textit{MainTC}. Then, customer creates a payment message based on \textit{MainTC}. In payment message customer determines a few refund addresses which are extended public keys of refundees, $((Pk_{r_1},ch_{r_1}),(Pk_{r_2},ch_{r_2}),\dots,(Pk_{r_k},ch_{r_n}))$, for $n$ refundees, and the amount of bitcoins each address should receive in case of canceling the order or overpayment. 
\item[\textbf{Payment ack.}] Merchant detects \textit{MainTC} and returns an acknowledgement message to customer.
\item[\textbf{Refund request.}]  Within a predetermined distance from payment request, customer can use the addresses provided in \textit{Payment.refund$\_$to} field to receive the refund.  In this case, merchant,
\begin{enumerate}
\item Splits each refund value (for different customers) to $k$ partition, for example, $v_1$ is divided to $v_{11}$, $v_{12}$, $\dots$ $v_{1k}$, and $v_2$ to $v_{21}$, $v_{22}$, $\dots$ $v_{2k}$, and so on.
\item Derives child refund keys $Pk^{\prime}_{r_{ij}}$, $\forall 1 \leq i \leq n$, where $n$ is the number of refundees, and $\forall 1 \leq j \leq k$. 
\item Derives child keys of customers $Pk^{\prime}_{c_{ij}}, \forall 1 \leq i \leq n+1$ and $\forall 1 \leq j \leq k$, according to the number of refundees $n$ each customer has.
\item  Generates refund transactions: In \textit{RefundTC1} each output (a refund chunk) is locked to a child key of refundee and customer, whereas in \textit{RefundTC2}, at each output, the refund chunk is just locked to the child key of customer. Each child key is masked using the private key of merchant related to that transaction. For example if the merchant public key is $Pk_m=m^{*}P$, the child key of customer is $Pk^{\prime}_c=c^{\prime}P$, then the masked child key  of customer will be $Pk^{\dprime}_{c}=Pk^{\prime}_{c}+H_{l}(m^{*}Pk^{\prime}_{c})P$.
\item  While mixing the outputs of multiple customers in each transaction Broadcasts the them to Bitcoin network.
\end{enumerate}

\end{description}

\end{document}